\documentclass[%
 reprint,
 amsmath,amssymb,
 aps,
]{revtex4-2}

\usepackage{ragged2e}
\usepackage{graphicx}
\usepackage{dcolumn}
\usepackage{bm}

\begin{document}

\preprint{APS/123-QED}


\title{Hydrodynamic Spin-Pairing and Active Polymerization of Oppositely Spinning Rotors}


\author{Mattan Gelvan}
\author{Artyom Chirko}
\author{Jonathan Kirpitch}
\author{Yahav Lavie}
\author{Noa Israel}
\author{Naomi Oppenheimer}
 \affiliation{Department of Physics and Astronomy and the Center for the Physics and Chemistry of Living Systems, Tel-Aviv University.}

\date{\today}

\begin{abstract}

Rotors are common in nature --- from rotating membrane-proteins to superfluid-vortices. Yet, little is known about the collective dynamics of heterogeneous populations of rotors.
Here, we show experimentally, numerically, and analytically that at small but finite inertia, a mixed population of oppositely spinning rotors spontaneously self-assembles into active chains, which we term gyromers. The gyromers are formed and stabilized by fluid motion and steric interactions alone. A detailed analysis of pair interaction shows that rotors with the same spin repel and orbit each other while opposite rotors spin-pair and propagate together as bound dimers. 
Rotor dimers interact with individual rotors, each other, and the boundaries to form chains. 
A minimal model predicts the formation of gyromers in numerical simulations and their possible subsequent folding into secondary structures of lattices and rings.
This inherently out-of-equilibrium polymerization process holds promise for engineering self-assembled metamaterials such as artificial proteins. 

\end{abstract}

\maketitle

Life is too complicated to be assembled manually \cite{schrodinger2012life}. From the nanoscopic starting point of the lipids that comprise the cell and the DNA molecules that hold the blueprints of each organism's design, to the final macroscale product of the organism itself, there are many length scales and billions of molecules that take part \cite{alberts2022molecular}. Life is also inherently out of equilibrium and forms far more complex structures than the crystalline structures expected for atomistic systems at low temperatures \cite{kittel2021introduction}. A prototypical example is proteins, which are self-assembled hierarchically on the molecular scale, first forming chains and then folding into secondary and tertiary structures which can function as intricate molecular machines \cite{doi1988theory}.

A growing body of research in recent years has been devoted to the study of self-assembly out-of-equilibrium. Yet, when the building blocks are isotropic, most of the resulting structures have been crystals, both in and out of equilibrium \cite{theurkauff2012dynamic, palacci2013living, briand2016crystallization, briand2016crystallization, mognetti2013living}. Breaking of symmetry is possible with the aid of an external field, such as electrostatic chains that align along field lines \cite{jennings1990electro, KIM2003149}, Janus particles with electrostatic imbalance \cite{yan2016reconfiguring}, active shakes oriented by an external field \cite{junot2023large, shoham2023hamiltonian}, magnetic dipoles in a perpendicular field \cite{PhysRevE.70.051502} or patchy colloids \cite{mcmullen2022self}. 

Despite the fundamental role of fluid dynamics in biological systems, hydrodynamic interactions have received limited attention in the study of active self-assembly. Notably, hydrodynamic structure formation by rotating objects has been garnering increasing interest. For example, synthetic rotors driven by a magnetic field \cite{grzybowski2000dynamic, yan2015rotating, soni2019odd, massana2021arrested} or light \cite{zion2022hydrodynamic}, living rotating matter \cite{drescher2009dancing, tan2022odd}, fluids or elastic materials with odd viscosity \cite{avron1998odd, banerjee2017odd, soni2019odd,bililign2022motile} or rolling particles,  \cite{bricard2013emergence, nguyen2014emergent, driscoll2017unstable}. These systems are ubiquitous in nature, ranging from the geological immensity of hurricanes to the biological intricacies of rotating proteins within cellular membranes and extending even to the quantum realm \cite{stockdale2020universal}. Historically, the behavior of rotors has been examined under conditions of negligible viscosity, dating back to Onsager \cite{onsager1949statistical}. More recently, studies have addressed settings of negligible inertia \cite{yeo2015collective, lushi2015periodic}, notably in biological contexts like ATP synthase proteins \cite{lenz2003membranes, oppenheimer2019rotating}. However, the intermediate Reynolds number regime, where both viscous and inertial forces play significant roles, remains relatively uncharted. Here, complexities arise due to the nonlinearity of the Navier-Stokes equations, leaving our understanding of this regime limited. Within this regime, research involving small spinning magnetic particles has demonstrated the formation of lattice structures \cite{grzybowski2000dynamic, grzybowski2001dynamic, goto2015purely}, highlighting the interesting dynamics that emerge in such systems.

\begin{figure*}[tbh]
\centering\includegraphics[width=\textwidth]{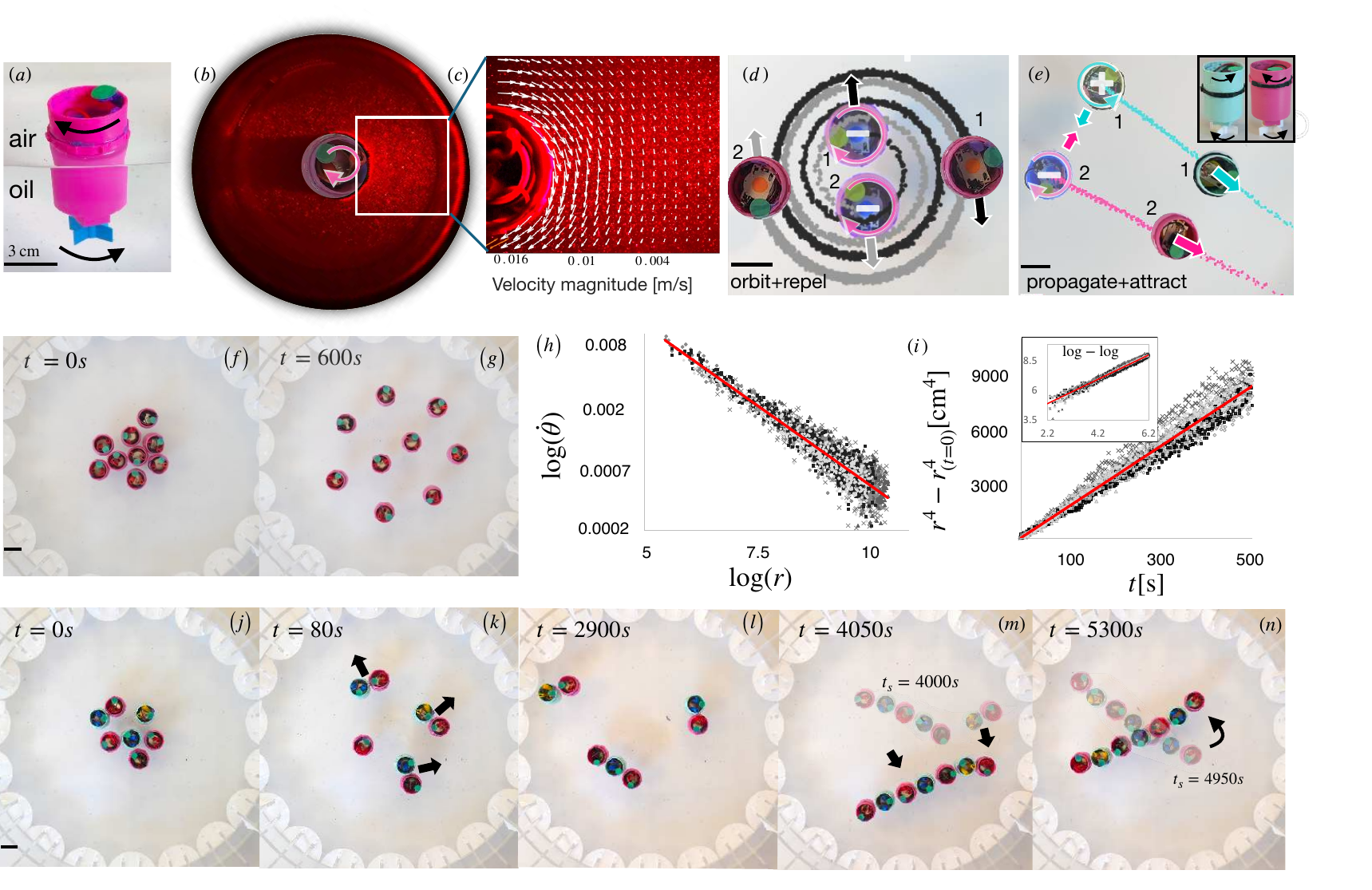}
    \caption{(a) A single rotor floating at the air-oil interface. The rotor is composed of a battery-driven motor in a 3D printed exoskeleton built of a cylindrical shell rotating in one direction and an oppositely rotating propeller. (b) PIV analysis of the flow induced by a single rotor shows a tangential flow field. (c) Velocity streamlines extracted from PIV images. (d) Two same-sign rotors spiral out (orbit around each other and repel). (e) Two counter-rotors self-propel and attract. 
    (h) Orbit angular velocity as a function of distance for two same-sign rotors in a log-log plot. Overlay of five experiments at different shades of gray. Solid red line marks the analytical prediction of $\dot{\theta} \propto 1/r^3$ (Eq.~\ref{eq:s_magnitude}). 
    (i) Distance to the fourth power as a function of time for the same five experiments of two same-sign rotors. Inset shows a log-log plot. The red solid line is a linear fit, verifying Eqs.~\ref{eq:Um} and~\ref{eq:m_magnitude}.
    (f)--(g) Nine same-sign rotors starting from a centered random configuration form an hexagonal lattice. Similar to results by Ref.~\cite{grzybowski2000dynamic} even though here no external field is present. (j--n) Snapshots of seven rotors (four pluses and three minuses) starting from a centered random configuration. The rotors quickly spin-pair and form dimers that self-propel. At longer times, gyromers composed of trimers, tetramers, and pentamers are formed until, finally, a single gyromer of all seven rotors is assembled. Once formed, the gyromer is stable and stays intact. The scale bar is 3 cm in all figures.
    }
    \label{fig1}
\end{figure*}

Common to all the aforementioned instances is that the units creating the self-organization spin in the same direction. A question arises: what unique properties will be observed in a mixture of binary rotors, where some spin clockwise while others counter-clockwise? At intermediate Reynolds, this is still unknown. Both in the limit of an ideal fluid with no viscosity and in the limit of purely viscous fluids the equations of motion have time reversal symmetry. Theoretical studies of binary rotors in these two limits have shown that the two populations separate and dense clusters of same-spin rotors are formed \cite{onsager1949statistical, yeo2015collective}. In fact, even completely dry opposite rotors separate \cite{nguyen2014emergent, scholz2018rotating}. It is yet unknown what happens in the intermediate regime which is qualitatively different since time reversal symmetry is no longer obeyed. 

Here, we show that when inertia is small but not negligible, the resulting structures are very different --- mixed-sign circular rotors self-assemble in a hierarchical manner, reminiscent of polymerization, but driven by fluid flow and steric interactions alone --- first, rotors spinning clockwise and counter clockwise attract, spin-pair and form bound dimers. Dimers then assemble into longer active chains. The chains, which we term gyromers (gyroscopic polymers), are stabilized by their activity (see Fig.~\ref{fig1}). To study the stability and origin of gyromer formation, we developed a test bed that enables controlling both the direction of rotation and the initial position of each individual rotor. Following Grzybowski, Stone, and Whitesides \cite{grzybowski2001dynamic}, we analytically describe the velocity field of numerous spinning bodies within an intermediate Reynolds number regime, where viscosity remains the dominant force, and inertia is a perturbation. We conduct numerical simulations that qualitatively reproduce the experimental results. We determine that gyromer stability is decreased with increcing concentration, as well as by the variance in angular velocities of individual rotors. 

\textbf{Experimental Setup}.
\begin{figure*}[t]
\centering\includegraphics[width=0.8\textwidth]{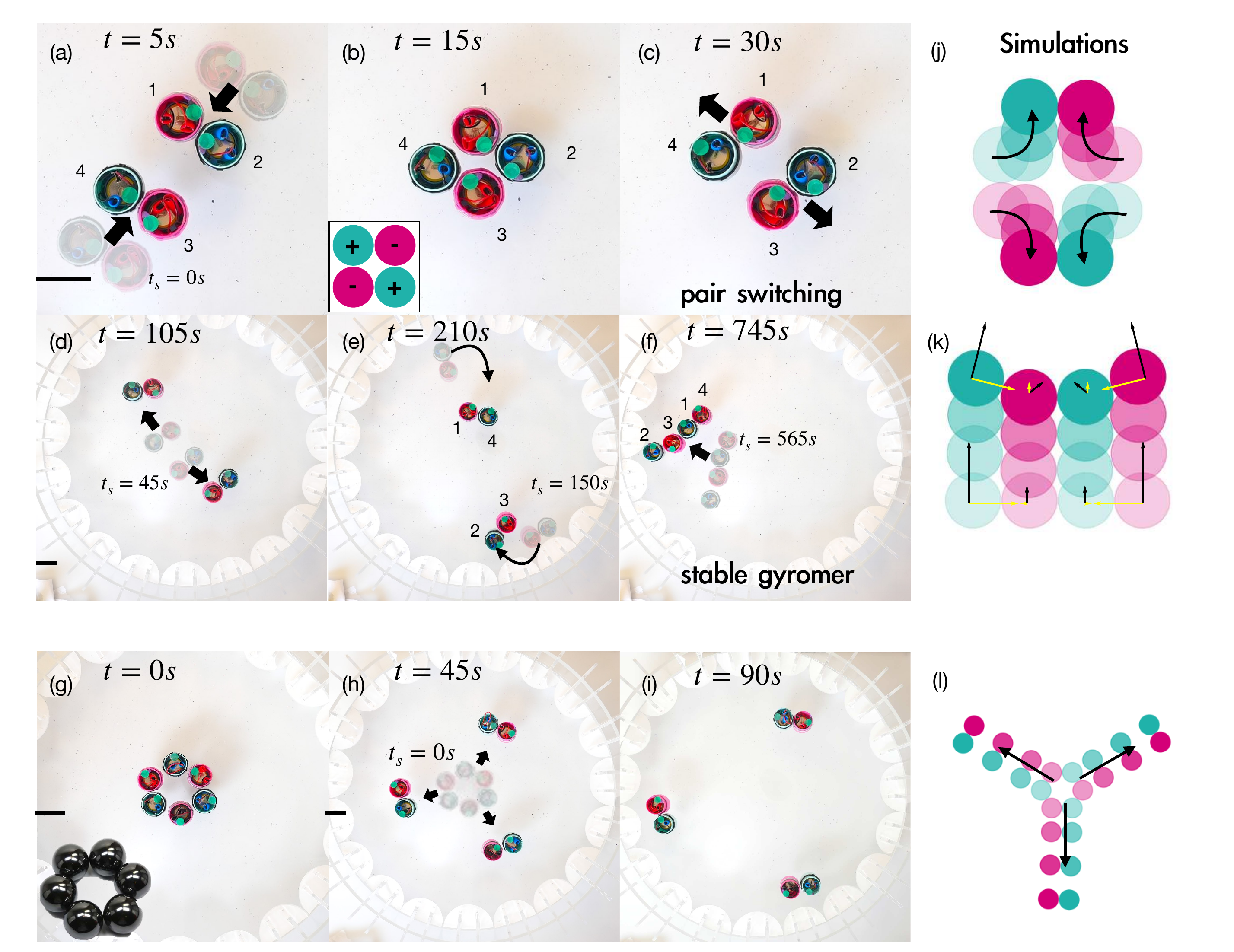}
\caption{(a) Initial position of two dimers advancing towards each other. (b) The four rotors collide and form a structure similar to an ionic lattice. (c) The lattice is unstable and immediately breaks into two new pairs of dimers, advancing in ninety degrees compared to the initial pairs. (d) The two dimers advance in space until reaching the boundary (e) The dimers turn due to the shape of the boundary  (see explanation in main text) and advance towards each other. (f) The dimers meet each other in a way that enables the formation of a 4-gyromer (tetramer) which self-propels in space. (g) Ring initial formation. The inset shows a static magnetic ring. (h) The ring breaks into three dimers that propagate forward (i) until they reach the boundary (later forming longer gyromers). (j) Simulation of two dimers colliding and switching pairs with $C_i = 0.8$. (k) Simulation of a 4-gyromer advancing through space ($C_i = 80$). Arrows are analytic solutions of the velocity field initially and at steady-state. Black arrows indicate the viscous part, and yellow the inertial part, giving a stabilizing component for an out-of-line deviation. (l) Simulations of an initial ring structure of six rotors with $C_i = 0.8$ break into three dimers as in the experiment. The scale bar is 4 cm in all figures}\label{fig2}
\end{figure*}
Rotors are built of a brushless motor (manufacture, 80 rpm in air) and a battery (LIR2477)  placed in a 3D-printed cylindrical shell (radius $a = 1.8$ cm, 7.5 cm height) that enables them to float when placed in a bath of silicone oil (diameter 60 cm, height 9.5 cm, viscosity $\mu=1 \,{\rm Pa}\cdot$s, density $\rho = 10^3 \, {\rm kg}/{\rm m}^3$). We use small belts positioned around the rotors to reduce friction with each other and with the bounding walls.  A 3D-printed propeller is attached to the bottom pin of the motor such that the two parts counter-rotate (see Fig.~\ref{fig1}A). The direction of rotation can be easily switched by changing the battery wiring. Particle Image Velocimetry (PIV) of the flow field was performed using Thorlabs-HNL050L laser and PIVlab software (Fig.~\ref{fig1}B). The resulting flow is tangential to the particle. 
The bath is continuously monitored from above by a Sony camera ($\alpha7s_3$, lens FE 4/24-105 G-OSS), capturing snapshots at one-second intervals. We track the trajectory of each rotor in Python using OpenCV (see SI). By marking each rotor with an off-centered dot, we can calculate its angular velocity. When placed in the oil bath, the rotors spin at 7 rpm ($\Omega \approx 0.7$ rad/s). The Reynolds number, a measure of inertial forces compared to viscous ones, is ${\rm Re} = \rho \Omega a^2/\mu \approx 0.2$. Even though the Reynolds number is small, inertia is not completely negligible. In fact, it changes the behavior of the system qualitatively. Two rotors at zero Reynolds number maintain a fixed distance, they do not draw nearer nor drift apart \cite{lenz2003membranes, yeo2015collective, oppenheimer2019rotating} --- same sign rotors orbit around each other and opposite sign rotors propagate. When inertia is included, it adds radial forces reminiscent of electrical charges --- same sign rotors orbit and also repel, tracing a growing spiral (Fig.~\ref{fig1}D); opposite sign rotors propagate and also attract (Fig.~\ref{fig1}E and Supplementary Video 1). The dynamics are predicted by the theoretical model described below. 

At higher numbers of rotors, the particles self-assemble into active chains (gyromers). Gyromers are constantly assembled and disassembled due to interaction with other rotors or with the boundaries. We commonly see pair-switching or companion stealing. Even-numbered gyromers self-propel in a direction perpendicular to their axis (see Fig.~\ref{fig2}F) with a speed that decreases with the number of monomers (an infinite gyromer will be stationary). Odd gyromers orbit around their center (see Fig.~\ref{fig1}N). Small changes in the spin of the rotors cause deviations from this ideal behavior. Gyromers are stable until encountering other rotors or the boundaries. Odd gyromers are stable for a longer duration since they are less prone to interact with the boundaries. As demonstrated in Fig.~\ref{fig1}, at a low concentration (7 rotors), all the rotors assemble into a single gyromer. Once formed, the gyromer remains stable for the lifetime of the rotors' batteries (a couple of hours). 

The boundary in our system has a critical role in facilitating the creation of self-assembled structures. Without it, assuming an infinite space, even-numbered gyromers can propagate further away. The boundary confines the rotors, enabling the formation and dissolution of assembled configurations. 
The boundary also introduces complexities. Initially, we designed either a circular container or a flower-shaped boundary \cite{Deseigne2010}, but rotors were hydrodynamically attracted to the boundary and tended to stay there \cite{blake1974fundamental}.
To eliminate attraction, we designed the boundary to be an inverse flower-like shape. This unique shape, designed by principles outlined in Refs.~\cite{arbel2024mechanical, Casiulis2024}, creates effective repulsion and brings rotors back into the pool.

\textbf{Theoretical Model}.
\begin{figure*}[tbh]
\includegraphics[width=1\textwidth]{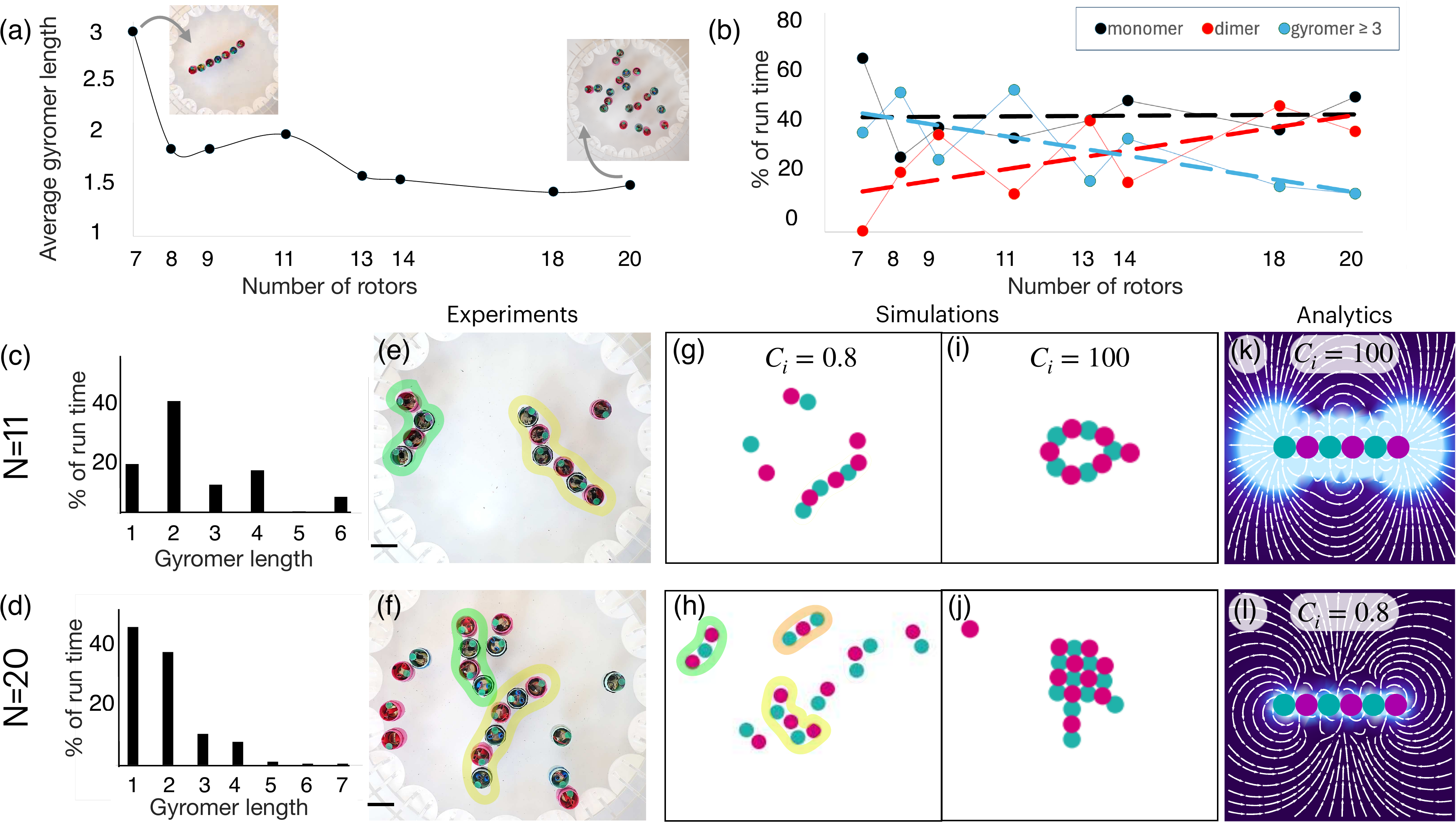}
\caption{\label{fig3}(a) Average chain length as a function of the number of rotors shows that at low concentrations, long gyromers are stable for longer periods. Two typical snapshots from experiments are shown. (b) Percentage of time monomers (black), dimers (red), and longer gyromers (blue) were observed in an experiment as a function of the number of rotors. Data is scattered but on average, the number of monomers does not depend on concentration, dimers increase with concentration and longer gyromers decrease. (c) and (d) Histograms of the percentage of chain length for 11 and 20 rotors. (e) and (f) Snapshots from experiments starting from random initial conditions with 11 (top) and 20 (bottom) rotors. Highlighted are the formed gyromers (the scale bars are 4 cm). (g) and (h) Simulations with inertial forces that are similar to the experimental ones show gyromer formations highlighted in colors. (i) and (j) Simulations with higher inertial forces initially form gyromers but, at later times, fold into tertiary structures of rings and lattices. (k) and (l) Analytic streamlines around a gyromer of six rotors showing flow lines that a positive rotor would be advected along. Two values of inertial constants are shown, $C_i = 100$ (top), where flow lines resemble a magnetic dipole and $C_i = 0.8$ (bottom) showing flows that resemble a torque dipole but with a component that breaks left-right and top-bottom symmetry around the middle of the chain.}
\end{figure*}
A single rotor far from boundaries stays static. When placed in a bath of other rotors it is advected by the flow created by all other rotors. 
In this region of the Reynolds number, the flow is governed mainly by viscous forces, but not only. We account for inertia by expanding the Navier-Stokes equations to a first order for a small but finite Re. In this limit, the equations are linear, and we can write ${\bf u} = \mathbf{u_s} + \mathbf{u_i}$, where $\mathbf{u_i}$ is the inertial correction and $\mathbf{u_s}$ is the part coming from the Stokes equations. The flow field of a single rotor in a viscous fluid is \cite{kim2013microhydrodynamics}
\begin{equation}
  \mathbf{u_s} \propto \frac{\mathbf{{\Omega} \times \hat{\mathbf{r}}}}{r^2},
  \label{eq:Us}
\end{equation}
where $\mathbf{\Omega}$ is the spin of the rotor.
Two identical, same-sign rotors in this viscous regime orbit around each other with an angular velocity, $\dot{\theta}$, given by 
\begin{equation}
\label{eq:omega}
\dot{\theta} \propto \Omega/r^3,
\end{equation}
while maintaining the initial distance between the rotors, $r$. When inertial effects are included, the second rotor adds a correction to the above flow, giving a lift force resembling the Magnus effect. Solving the Naiver-Stokes equations under a small but finite Reynolds number for a spinning disk in the shear rate created by a second, faraway disk gives the leading order correction to a viscous flow, which is radial and has the scaling~\cite{schonberg1989inertial, grzybowski2000dynamic, grzybowski2001dynamic}
\begin{equation}
    \mathbf{u_i} \propto{} \frac{\rho \Omega_1\Omega_2l_c^7}{r^3}\hat{r}.
    \label{eq:Um}
\end{equation}
In this low-but-finite-Reynolds regime linearity is maintained and the velocity of the $j$th particle, ${\bf u}_j$, is given by to the total flow created by all other particles.  In complex notation $z_j = x_j +iy_j$, we can use Eqs.~\ref{eq:Us} and \ref{eq:Um} and write
\begin{equation}
    \ \dot{z_{j}} = \sum_k iC_s \Omega_k \frac{z_j - z_k}{|z_j - z_k|^s} + C_i \Omega_j \Omega_k \frac{z_j - z_k}{|z_j -z_k|^m},
    \label{eq:zdot_field}
\end{equation}
where $\Omega_j$ $j$th rotor own spin, and $C_s$ and $C_i$ are constants. For completeness we have kept the power laws $s$ and $m$ general. Using the scaling in Eq.~\ref{eq:Us} and Eq.~\ref{eq:Um} gives $s = 3$ and $m = 4$, which we verify experimentally in Fig.~\ref{fig1}M and Fig.~\ref{fig1}N. 

Two key assumptions were made here: First, we assumed that the rotors act as rotating disks. The use of self-spinning brushless motors in an infinite fluid results in a torque dipole along the Z-axis, as illustrated in Figure \ref{fig1}A. It is, therefore, not immediately apparent that the magnitude of the angular velocity of a pair of rotors should indeed scale as Eq.~\ref{eq:omega}. However, the vicinity of the propeller to the bottom of the container dampens its decay \cite{blake1974fundamental}, as we have verified from PIV measurements (see SI). Therefore, the dynamics are governed by the upper cylinder. Second, Eq.~\ref{eq:Um} is the far distance first approximation of the flow, though in practice, this is often not the case. Certainly, this is not correct for nearest neighbors in a gyromer. 
To verify the validity of both assumptions when rotors are distant, we analyze the dynamics of two same-sign rotors. 

\begin{figure}[tbh]
\includegraphics[width=0.9\linewidth]{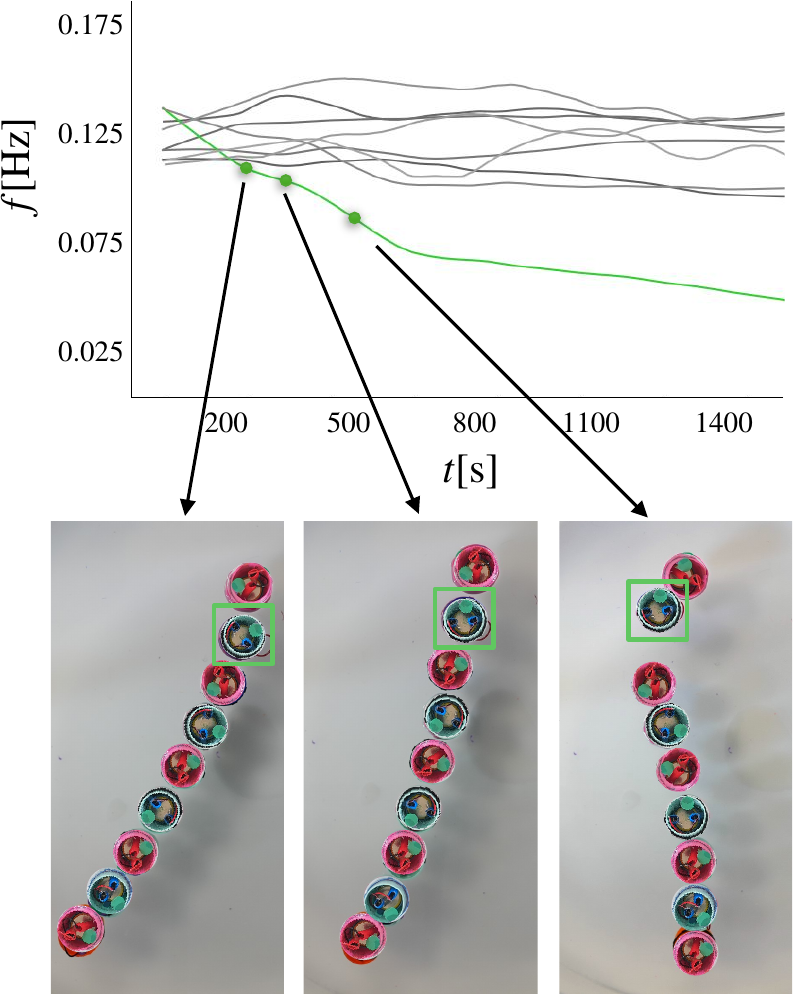}
\caption{
The moment of breakage of a 9-gyromer. The gyromer broke into a dimer and a 7-gyromer. Analysis of the spin of the rotors (frequency, $f$) shows that all spins are within $\pm 10\%$ of each other until one of the rotors (marked in green) is starting to fail. When its spin is decreased by more than twenty percent, the gyromer starts to disintegrate. Snapshots at three points in time are shown at the bottom.}
\label{fig4}
\end{figure}

Solving analytically Eq.~\ref{eq:zdot_field} for two rotors gives
\begin{align}
    &r^m = 2mC_i\gamma \Omega^2 t + A
    \label{eq:m_magnitude} \\
    &\dot{\theta}= \frac{(\gamma +1)C_s\Omega}{r^s},
    \label{eq:s_magnitude}
\end{align}
where we have chosen $\Omega_1 = \Omega$ and $\Omega_2 = \gamma \Omega$. Note that the distance between rotors only depends on the inertial term and that it grows for same-sign rotors and decreases for opposite-sign rotors. Conversely, the orbiting angular velocity over time only depends on the viscous term and is inversely proportional to the distance between the rotors to the $s$th power. 
We experimentally tested the dynamics of several two same-sign rotors with $\gamma \approx 1$ (Supplementary Video 1). Figure~\ref{fig1}H shows the angular velocity of the pair with a power law of $s = 3$, as predicted by Eq.~\ref{eq:omega}. That is, $u_s \propto 1/r^2$. Figure~\ref{fig1}I shows that the separation distance to the forth power is linear with time, as predicted by Eq.~\ref{eq:m_magnitude} ($m = 4$), that is $u_i \propto 1/r^3$.


As we advance to study the interaction involving multiple rotors, we will compare our system to two well-known equilibrium systems. The first is charged particles, which also repel (same sign) or attract (opposite sign) and are prone to form ionic crystals \cite{atkins2023atkins}. The second system that bears a resemblance to ours is that of magnetic dipoles, which are known to form chains and rings \cite{jackson2021classical}. For more than two rotors, solving Eq.~\ref{eq:zdot_field} analytically becomes challenging. Instead, we will leverage the insights gained from the two-rotor scenario and introduce the four-rotor and six-rotor interaction experimentally, numerically and use linear stability analysis.

\textbf{Four Rotors}. Two rotors of opposite sign attract and propagate in space. In fact, in our experiments, spin pairing is very common. Once the rotors make contact, they advance in space as a bound dimer. The direction of their motion is perpendicular to the dimer orientation. To study the interaction between four rotors, we initilized two rotor-dimers such that they advance toward each other, as shown in Fig.~\ref{fig2}A. The two pairs collide in a formation resembling an ionic crystal of plus and minus charges (Fig.~\ref{fig2}B). However, unlike static ionic crystals, in our system, these lattice formations are never stable and they quickly break into two new pairs of rotor-dimers (Fig.~\ref{fig2}C). This effect is purely due to the dynamic nature of the interactions. Unlike static lattices, the forces governing the system include not only attraction and repulsion between particles but also angular forces that break the symmetry causing the crystal to disperse. The pairs then reorient due to interaction with each other or with the walls (Fig.~\ref{fig2}E) until reforming into a 4-rotor gyromer (Fig.~\ref{fig2}F). Once formed, the gyromer is stable for a long time ($\sim$ half an hour, see Supplementary Video 2). Simulations of four rotors were performed by propagating Eq.~\ref{eq:zdot_field} using a fifth-order Runge-Kutta scheme in Python (see SI). The simulations reproduce the experimental results taking $C_s = 1$, and $C_i$ is either 0.8 (Fig.~\ref{fig2}F and~\ref{fig2}L) or 80 (Fig.~\ref{fig2}K). The arrows overlaying Fig.~\ref{fig2}K are the analytical solution for the velocities in the initial and steady-state configuration. Black arrows represent the viscous forces, and yellow arrows the inertial ones. Notice how the viscous forces drive the outer and inner pairs apart, but once out of line, the inertial forces have a downward component that stabilizes the gyromer.

\textbf{Ring Formation}.
Since rotor-dimers are a common and relatively stable feature of our system we wanted to compare it to another system that forms chains --- magnetic dipoles. To that end, we examined an initial ring formation, as depicted in Fig.~\ref{fig2}G-i and Supplementary Video 3. The ring quickly explodes into smaller ``building blocks" of rotor-dimers. Subsequently, some of these dimers attract and eventually create larger gyromers. This behavior is starkly different from static magnetic dipolar systems, where ring formations remain stable (as seen in the inset of Fig.~\ref{fig2}G). 


\textbf{The Self Assembly of multiple Rotors}.
Up to this point, we have discussed simple and engineered cases of rotors. We now seek to discover general and statistical properties of gyromer formation, starting from random initial conditions. With increasing concentration, the availability of monomers increases, and longer chains form more rapidly but also dissociate more quickly. It is, therefore, unclear a priori what concentrations will produce longer gyromers. We conducted experiments with increasing numbers of rotors, and analyzed gyromer lengths in each frame. We observe that with increasing concentration, the average gyromer length decreases (see Fig.~\ref{fig3}). We also studied histograms of gyromer lengths. At all concentrations, the probability of having dimers is the highest. This probability increases with concentration. The number of single rotors (monomers) is constant with concentration, and the number of gyromers ($N\geq 3$) decreases with concentration.
See Fig.~\ref{fig3}. 

Comparing experimental results to simulations at different concentrations shows similarities and differences. Using inertial values that are similar to the experimental ones, we observe that gyromers in the simulation are less stable and more quickly dissociate into dimers and trimers (Supplementary Video 5). There can be several reasons for deviations from the simplified model --- (a) at short distances, our theoretical model is no longer expected to hold. Indeed, experimentally, we observe deviations even for a pair of opposite rotors as they approach distances of $\approx 1.5 a$, with $a$ being the radius of the rotor (see SI). (b) There may be deviations from simple pair interactions such that superposition is not possible. (c) Other interactions, though weaker, come into play, such as attraction due to capillary forces (the Cheerios effect, see SI). 
Running simulations with much higher inertial interactions, $C_i =100$, shows that gyromers are quickly assembled and are more stable, but at later times fold to a tertiary structure of rings and square lattices, which remain stable for the rest of the simulation (see Fig.~\ref{fig3}i and Fig.~\ref{fig3}j).


Lastly, another key ingredient to the stability of the gyromers is the spin of individual rotors. We analyzed gyromers consisting of 7--11 monomers, which were stable for up to a couple of hours when no free monomers were present. A question arises: what makes the gyromer break after such a long time of stability? (Supplementary Video 6).
From the analysis of the rotor's spin, it can be seen that there are always deviations in frequencies of up to $10\%$ (see Fig.~\ref{fig4}). In fact, in simulations, we see that adding noise to the spins increases gyromer stability. However, once the difference is greater than around $20\%$, gyromers dissociate. In experiments, this happens when one of the batteries starts to fail. A typical instance of gyromer breakage is shown in Fig.~\ref{fig4} along with the failure of one of the rotors, marked in green. 

\textbf{Conclusions}.
We have revealed the novel dynamic self-assembly of mixed-sign rotors, demonstrating that due to the combination of viscous and inertial forces, oppositely spinning elements attract and propagate to form active structures termed gyromers. Unlike homogeneous rotor systems that lead to the formation of hexagonal lattices, our mixed-sign rotors orchestrate themselves into linear assemblies driven purely by hydrodynamic and steric interactions with no external fields. Without perturbations from the boundaries or other rotors, the chains are stable for long durations, but due to vigorous activity in the bath at high concentrations, chains constantly form and dissociate. Simulations indicate that when increasing inertial forces, gyromers fold into tertiary structures of square lattices and rings which remain stable. 
These findings not only enhance our understanding of active matter systems but also lay a foundational step towards engineering advanced materials and devices harnessing the self-organizing principles observed in natural and synthetic active systems, presenting opportunities for future explorations of the complex interplays of force, motion, and structure.

\begin{acknowledgments}
We wish to thank Matan Yah Ben Zion and Yoav Lahini for their helpful suggestions and discussions. This research was supported by the Israel Science Foundation (grant No. 1752/20).
\end{acknowledgments}

\appendix

\bibliography{gyromerBib}

\end{document}